# Bi$_2$Se$_3$ Growth on (001) GaAs Substrates for Terahertz Integrated Systems

Yongchen Liu[1], Wilder Acuna[1], Huairuo Zhang[2,3], Dai Q. Ho[1], Ruiqi Hu[1], Zhengtianye Wang[1], Anderson Janotti[1], Garnett Bryant[4], Albert V. Davydov[2], Joshua M. O. Zide[1], and Stephanie Law[1,*]

[1] Department of Materials Science and Engineering, University of Delaware, Newark DE 19716 USA

[2] Materials Science and Engineering Division, National Institute of Standards and Technology, Gaithersburg, MD 20899 USA

[4] Nanoscale Device Characterization Division, National Institute of Standards and Technology, Gaithersburg, MD 20899 USA

[3] Theiss Research, Inc., La Jolla, CA 92037 USA

[*]E-mail: slaw@udel.edu

Keywords: Bi$_2$Se$_3$, van der Waals materials, scanning transmission electron microscope, molecular beam epitaxy, GaAs/AlAs superlattice

**Abstract**

Terahertz (THz) technologies have been of interest for many years due to the variety of applications including gas sensing, nonionizing imaging of biological systems, security and defense, etc. To date, scientists have used different classes of materials to perform different THz functions. However, to assemble an on-chip THz integrated system, we must understand how to integrate these different materials. Here, we explore the growth of Bi$_2$Se$_3$, a topological insulator (TI) material that could serve as a plasmonic waveguide in THz integrated devices, on technologically-important GaAs (001) substrates. We explore surface treatments and find that atomically smooth GaAs surface is critical to achieving high-quality Bi$_2$Se$_3$ films despite the relatively weak film/substrate interaction. Calculations indicate that the Bi$_2$Se$_3$/GaAs interface is likely selenium-terminated and shows no evidence of chemical bonding between the Bi$_2$Se$_3$ and the substrate. These results are a guide for integrating van der Waals materials with conventional semiconductor substrates and serve as the first steps toward achieving an on-chip THz integrated system.

**Introduction**

Terahertz (THz) technologies, which span the frequency gap between the infrared and microwave ranges, have been of increasing interest. THz devices enable a plethora of functionalities. For imaging and sensing, the THz photon energy is low enough to be nonionizing and thus it is biologically safe compared to imaging with x-rays.[1] Water molecules and most metals absorb THz radiation, providing a way to perform non-destructive security screening.[2] Another important application of THz light is materials characterization, since many materials can be identified by their THz optical responses.[3] Other applications involve high-data-rate wireless communication due to the large absolute bandwidth in this region, pharmaceutical quality control, and food safety analysis. [4-6]

To realize these applications, a THz integrated system that mimics existing optical photonic integrated circuits needs to be fabricated. However, this is hindered by the lack of necessary THz components, including photon sources, THz waveguides, modulators, and detectors. Recently, there has been significant effort devoted to creating such components.[7-9] However, despite many successes in creating individual components, there has been relatively little attention paid to integrating these separate components into one on-chip system. In this manuscript, we have explored the integration of a topological insulator (TI) with a technologically relevant semiconductor substrate.

Topological insulator (TI) materials are one of the prospective candidates for THz waveguides or plasmonic components.[10-13] TIs have an inverted band structure, with a bulk band gap crossed by linearly-dispersing surface states. These surface states house two-dimensional, massless, spin-polarized electrons which are of interest for a wide array of applications. Among other things, these electrons can host Dirac plasmons in the THz spectral range; these plasmons have been shown to have large mode indices and long lifetimes which makes them good candidates for plasmon-enhanced gas sensing or plasmonic waveguides.[10,11,14] However, before we can create devices from TI plasmonic structures, we must first learn how to integrate them with existing THz technology. TI materials include $Bi_2Se_3$, $Bi_2Te_3$, and $Sb_2Te_3$. Among these materials,

Bi$_2$Se$_3$ is widely used as it is relatively easy to synthesize and has a larger band gap than the other TI materials. [15] For these reasons, we chose to use Bi$_2$Se$_3$ as the TI material exemplar in this manuscript.

To date, most scientists have used inert substrates like sapphire for van der Waals materials growth. [16,17] These substrates are a natural choice since they have no dangling bonds and so are clearly compatible with vdW epitaxy. However, if we want to synthesize an on-chip system that includes vdW materials, we must understand how to grow these materials on traditional semiconductor substrates. There has been some previous work on the growth of Bi$_2$Se$_3$ on GaAs (111) since both materials have hexagonal in-plane symmetry. However, the GaAs (001) substrate has significant technological advantages for materials integration. Specifically, the growth of device structures on GaAs (001) is far more mature, with larger substrates readily available commercially. This permits the growth of complex III-V-based structures that can interact with the TI material. For THz generation and detection, materials such as low-temperature grown GaAs (LT-GaAs), ion-implanted GaAs, and ErAs nanoparticles incorporated in GaAs are most often grown on GaAs (001).[18,19] While LT-GaAs is the most commonly used semiconductor for THz generation, ErAs:GaAs is also of interest as it has shown an improvement in THz detection.[19] We are therefore interested in the growth of a TI on GaAs (001) where, for example, the TI can serve as a waveguide for THz radiation that can be emitted or detected by III/V structures. More generally, this interface work allows the possibility of integration of III-V devices with TI materials, such as back-gating with III/V materials to modulate TI carrier density.

Previous work regarding the MBE growth of Bi$_2$Se$_3$ on GaAs and related substrates has been done by a variety of groups. Some of this work has focused on the growth Bi$_2$Se$_3$ (0001)$_H$ on GaAs (111) substrates since these substrates have an in-plane hexagonal symmetry similar to that of Bi$_2$Se$_3$ [20-23]. Additional work has investigated the growth of Bi$_2$Se$_3$ (0001)$_H$ on GaAs (001) [15,24,25]. Finally, it has also been demonstrated that Bi$_2$Se$_3$ ($\bar{1}$015)$_H$ can be grown on GaAs(001) [14,26]. However, these papers generally focused on overall Bi$_2$Se$_3$ film quality rather than on understanding and improving the interface between the $Bi_2Se_3$ film and the technologically-important GaAs (001) substrate. A sharp interface is critical both to good film growth and to the creation of $Bi_2Se_3$/GaAs (001) devices. In addition, understanding the chemical makeup of the interface is important for accurate device modeling. In this paper, we concentrate on improving and characterizing and the $Bi_2Se_3$/GaAs (001) interface, understanding its influence on the $Bi_2Se_3$ film properties, and modeling the chemical nature of the interface.

To understand how best to grow $Bi_2Se_3$ on GaAs (001) using molecular beam epitaxy (MBE), we grew three samples with three different substrate preparations using semi-insulating single-side polished GaAs (001) wafers. The first sample had the GaAs native oxide on the surface, which was thermally desorbed under a selenium overpressure in the chalcogenide MBE. We will refer to this as the Se-desorbed sample. The second sample had the native oxide thermally desorbed under an arsenic overpressure in the III-V MBE before being transferred to the chalcogenide MBE via an ultra-high vacuum (UHV) transfer tube. We will refer to this as the As-desorbed sample. Finally, the last sample had the native oxide desorbed under an arsenic overpressure in the III-V MBE followed by deposition of a 100 nm AlAs/GaAs superlattice (SL) with a 10 nm period and a 50 nm GaAs buffer. The sample was then transferred to the chalcogenide MBE via the UHV transfer tube. We will refer to this as the SL sample. While previous results for growth on GaAs (001) demonstrated the effectiveness of a GaAs homoepitaxial layer for overall material quality [24,25], the SL structure was chosen (1) to result in an atomically-smooth semiconductor surface more quickly than a homoepitaxial GaAs layer[27-30], (2) to demonstrates the capabilities of incorporating this interface near the III-V heterostructures needed for future active devices, and (3) to assist with atom identification in transmission electron microscopy measurements. After entering the chalcogenide MBE, all substrates had 50 nm of $Bi_2Se_3$ deposited on them under identical conditions (for details, see Methods).

**Results and Discussion**

Atomic resolution high angle annular dark-field scanning transmission electron microscopy (HAADF-STEM) was performed to characterize the interface, which is a critical concern for heteroepitaxial systems. **Figure 1(a)** and (b) show typical images of the rough interface of $Bi_2Se_3$/GaAs in the Se-desorbed and As-desorbed samples, respectively. Both $(0001)_H$ and $(\bar{1}015)_H$ in-plane oriented $Bi_2Se_3$ domains were observed in these samples. The subscripts H, R, and C in this work indicate the hexagonal, rhombohedral, and cubic index system. $(\bar{1}015)_H$ is equivalent to $(015)_H$ and $(221)_R$ in $Bi_2Se_3$.[15,31,32] The $Bi_2Se_3$ vdW gaps are oriented parallel to the GaAs $(001)_C$ surface for the $(0001)_H$ orientation growth and oblique to the surface for the $(\bar{1}015)_H$ orientation growth. As shown in Figure 1(c), the $(0001)_H$-oriented $[1\bar{1}00]_H$ and $[\bar{1}2\bar{1}0]_H$ $Bi_2Se_3$ domains appeared on flatter regions of the GaAs $(001)_C$ substrate in the As-desorbed sample, while the $(\bar{1}015)_H$-oriented growth of the $[\bar{1}2\bar{1}0]_H$ $Bi_2Se_3$ domains occurred on a multi-faceted rougher area. The two intersecting white lines in Figure 1(c) drawn along the local $(\bar{1}015)_H$ crystallographic planes suggest the $(\bar{1}015)_H$ orientation was initiated by the local facets of the rough GaAs substrate in the As-desorbed sample.[20] A wide range of angles (49º - 58º) between the vdW gaps and the surface were observed in this case. In contrast, Figure 1(d) demonstrates a large-scale smooth interface over several micrometers of $Bi_2Se_3$/GaAs grown on the SL sample. By referencing the atomically-resolved AlAs dumbbell structure from the SL (Figures 1(e) and 1(g)), the atomically-sharp interface of $Bi_2Se_3$/GaAs was identified as terminated on the As-site (Figures 1(f) and 1(h)). Energy dispersive spectroscopy (EDS) was further performed to analyze the elemental interdiffusion across the interface of $Bi_2Se_3$/GaAs in the SL sample (Figure 1(i)). As shown in Figure 1(j), both Se and Bi diffuse into the GaAs substrate up to 2.1 nm, with a stronger diffusion of Se. In the diffusion zone of the substrate, the concerted reduction of Ga and As signals suggests that As and Ga sites were partially replaced by Se and Bi, respectively, leaving some Ga sites vacant. It should be noted that it is technically challenging to differentiate the single As and Se atoms at the $Bi_2Se_3$/GaAs interface by HAADF-STEM and EDS. Considering the selenium atmosphere in the growth chamber and the strong diffusion activity of Se, the final As-site at the interface of $Bi_2Se_3$/GaAs could be occupied by Se atoms, as indicated in Figures 1(f) and 1(h) with a notation $Se_{As}$. This is consistent with theoretical predictions discussed later in the paper which indicate that a selenium-terminated interface is more energetically favorable.

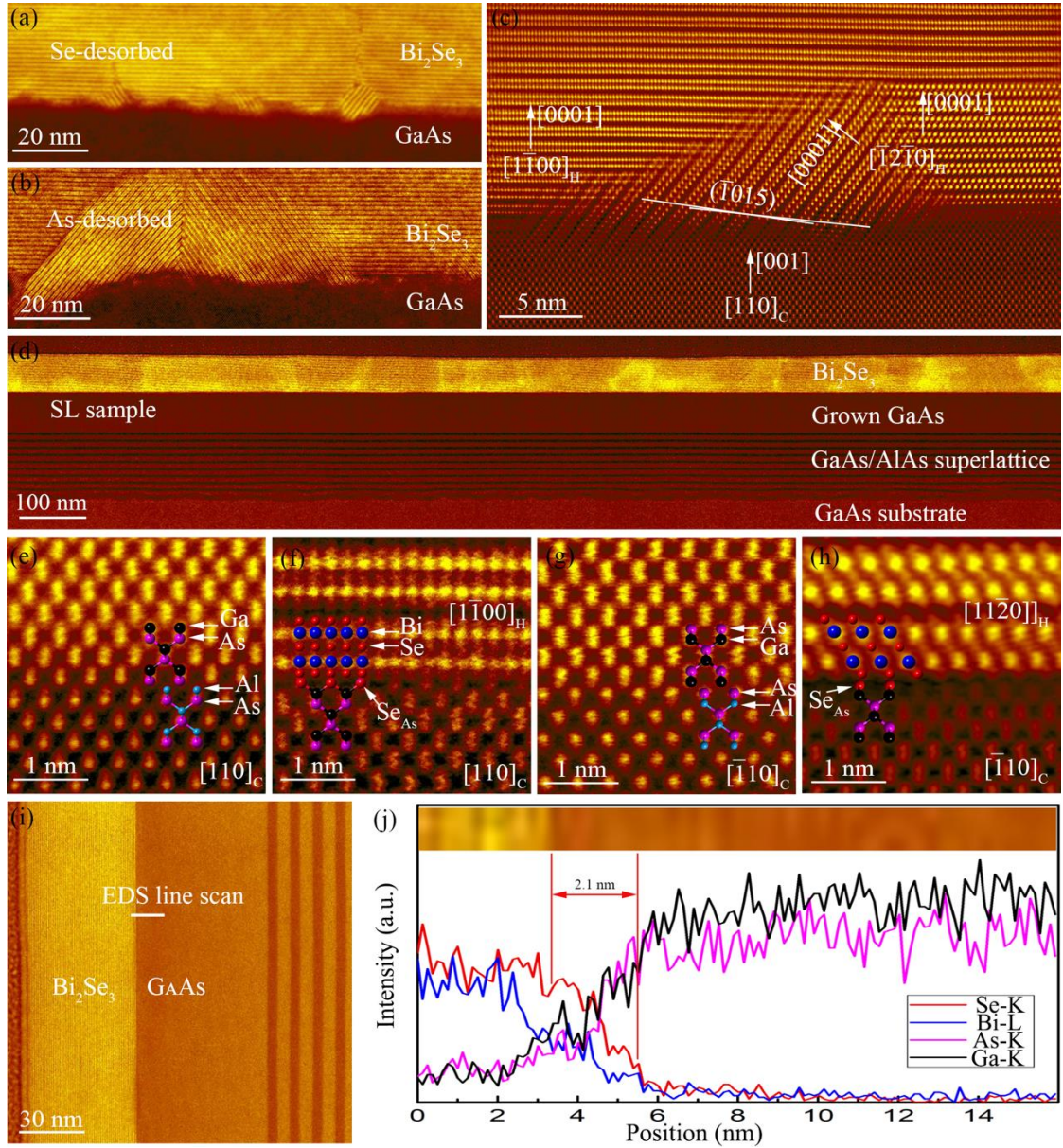

**Figure 1**. Cross-sectional HAADF-STEM characterization. (a,b) HAADF-STEM images taken along the GaAs $[110]_C$ zone-axis showing the rough interface of $Bi_2Se_3$/GaAs in the Se-desorbed and As-desorbed samples, respectively. (c) Zoomed-in atomic resolution image of the As-desorbed sample showing $(0001)_H$ oriented growth of $[1\bar{1}00]_H$ and $[\bar{1}2\bar{1}0]_H$ $Bi_2Se_3$ domains on a flat area of the GaAs $(001)_C$ substrate, and $(\bar{1}015)_H$ oriented growth of $[\bar{1}2\bar{1}0]_H$ $Bi_2Se_3$ domain on a rough multi-faceted area of the GaAs substrate. (d) Large-scale image demonstrating the flat interface of $Bi_2Se_3$/GaAs in the SL sample. (e-h) Zoomed-in images of the SL sample overlaid with atomic models showing the dumbbell atomic structure of the GaAs/AlAs superlattice in (e, g), and the atomically smooth interface of $Bi_2Se_3$/GaAs terminated on the Se replaced As-site (f, h) in the GaAs surface along the $[110]_c$ and $[\bar{1}10]_c$ directions,

respectively. (i) Rotated image showing the STEM-EDS line-scan analysis region in the SL sample. (j) Line-profiles of chemical compositions across the $Bi_2Se_3$/GaAs interface in the SL sample. False colors are added in the images to aid the eye.

We further characterize these samples using atomic force microscopy (AFM) and x-ray diffraction (XRD) measurements, shown in **Figure 2**. All three samples show similar surface morphology as measured by AFM in Figure 2(a-c). We observe triangular domains with a terraced morphology, as is common for $Bi_2Se_3$ films. The triangular morphology is caused by the symmetry of the unit cell while the terracing is caused by the vdW nature of the film, which makes it energetically possible to start a new layer before finishing the previous layer.[33,34] However, the SL sample has a lower root mean square (RMS) roughness at 1.29 nm than the other two (1.82 nm for the As-desorbed sample and 1.57 nm for the Se-desorbed sample). We attribute the reduced roughness to the smoother interface between the $Bi_2Se_3$ and the GaAs, as shown in Figure 1. The XRD data for the films are shown in Figure 2(d) and (e). In Figure 2(d), the scans are taken with the x-ray beam parallel to the GaAs $[0\bar{1}\bar{1}]_C$ axis, while in Figure 2(e), the scans are taken with the direction of the x-ray beam perpendicular to the GaAs $[0\bar{1}\bar{1}]_C$ axis. Scans were taken in both directions since previous work indicated an anisotropy in the growth of the $(\bar{1}015)_H$ orientation of $Bi_2Se_3$ due to the diffusion anisotropy of bismuth on the surface of GaAs [20]. The scans for the SL, Se-desorbed, and As-desorbed samples are shown in purple, green, and blue, respectively. Red and black spectra show the scans for the GaAs/AlAs SL and GaAs substrate. In the XRD scans, we see peaks near 18º, 28°, 38º and 48°, which are the $(0006)_H$, $(0009)_H$, $(00\underline{12})_H$ and $(00\underline{15})_H$ reflections from the $Bi_2Se_3$ film, respectively. The peak indicated by a dashed line that appears near 29.5º can be attributed to a reflection from the unwanted $(\bar{1}015)_H$ orientation of $Bi_2Se_3$ film [16, 26]. We see this peak in the Se-desorbed and As-desorbed samples only; differences in peak intensity for the scans parallel (Figure 2(d)) and perpendicular (Figure 2(e)) to the GaAs $[0\bar{1}\bar{1}]_C$ axis are caused by the aforementioned anisotropy of the growth of the $(\bar{1}015)_H$ orientation. We do not observe a $(\bar{1}015)_H$ peak in the SL sample, consistent with the smooth interface for the SL sample shown in Figure 1. Other peaks in these spectra are from GaAs substrate or the GaAs/AlAs SL and can be found in the corresponding spectra. The triple-peak near 32º in the red and purple scans comprises the zeroth-order SL peak as well as two satellite peaks. X-ray pole scans were also performed to quantify the degree of twinning in the $Bi_2Se_3$ films and can be found in the Supplementary Information; no significant difference was observed among the samples as the different surface treatments were only intended to control the surface roughness, not the surface energy.[31]

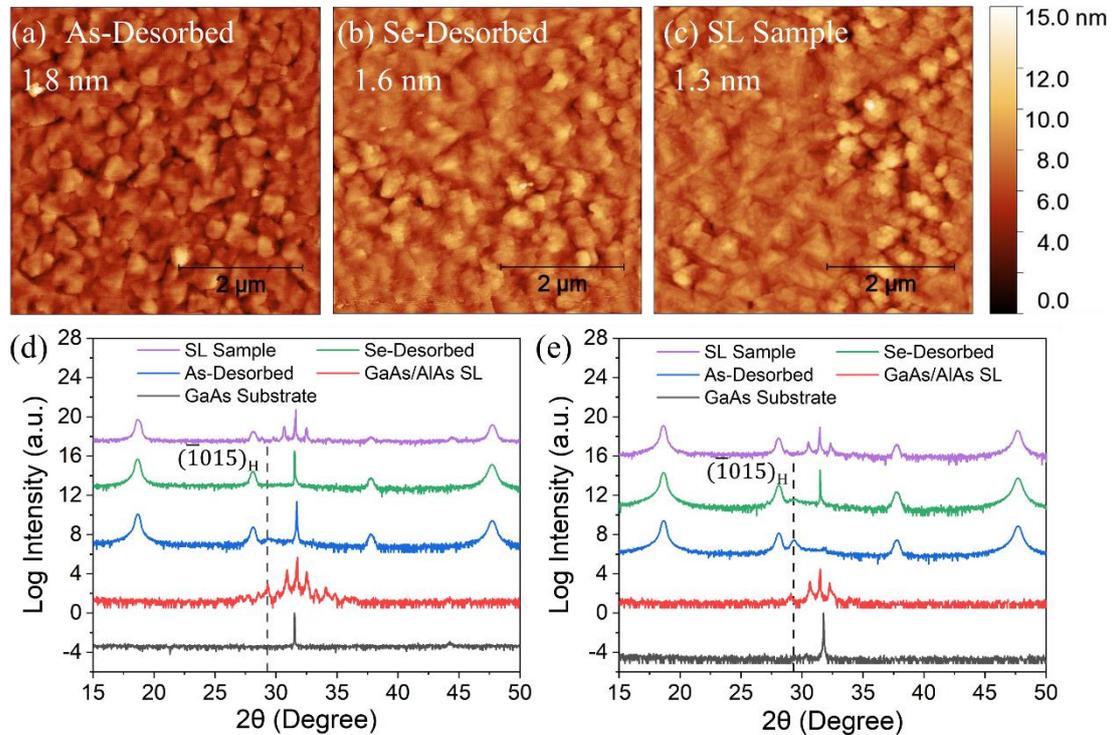

**Figure 2.** Atomic force microscopy (AFM) images of the As-desorbed sample (a), the Se-desorbed sample (b), and the SL sample (c). RMS roughness is indicated on each image. X-ray diffraction scans of the three films and the GaAs substrate with the incident direction of the x-ray beam aligned parallel (d) and perpendicular (e) to the GaAs $[0\bar{1}\bar{1}]_C$ axis. Vertical dashed lines in (d) and (e) indicate the $(\bar{1}015)_H$ reflection position.

Finally, we characterized the transport properties of these three films using room-temperature Hall effect measurements in the van der Pauw configuration.[35] This is a useful metric since most applications of TI materials rely on transport through the surface states. A perfect TI film with the Fermi energy in the bulk band gap would show a low carrier density and high mobility since all the current would be carried by the surface states. Typical TI films show relatively high carrier densities and low mobilities at room temperature. The high carrier densities can be caused by defects at the film/substrate interface, by point defects throughout the film, and/or by band bending accumulation layers at the substrate and/or air interfaces.[36] The Se-desorbed sample has mobility of 247±0.38 cm$^2$/Vs and a carrier density of 4.60±0.01×10$^{13}$ cm$^{-2}$. The As-desorbed sample has mobility of 466±1.18 cm$^2$/Vs and a carrier density of 3.74±0.03×10$^{13}$ cm$^{-2}$. The SL sample has mobility of 525±0.98 cm$^2$/Vs and a carrier density of 3.95±0.01×10$^{13}$ cm$^{-2}$. These transport measurements support the assessment that the SL sample has the highest structural quality, as indicated by its higher mobility. Interestingly, the mobility of the As-desorbed sample is higher than that of the Se-desorbed sample and the As-desorbed sample has a lower carrier density. This seems to imply that the As-desorbed sample has a better structural quality than the Se-desorbed sample despite having a higher RMS roughness and a similar XRD scan. This seeming inconsistency could be caused by a difference in band-bending in the TI at the Bi$_2$Se$_3$/GaAs interface. A larger upward band-bending in the As-desorbed sample, for example, could lead to a lower carrier density and a higher mobility since more of the current would be carried through the bulk of the film, away from the defective interface. Despite the atomically-flat interface, the SL film still has an overall relatively low mobility and high carrier density when compared to TI films grown on vdW buffers.[37-39] Using a thin vdW buffer layer to improve the film mobility for future devices is reasonable. Generally, vdW buffer layers are only a few nanometers thick, much thinner than the wavelength of THz light. If we consider using the Bi$_2$Se$_3$ layer as a THz waveguide, as described previously, a few nanometer vdW buffer layer grown from a wide band gap material like In$_2$Se$_3$ is unlikely to negatively impact device performance while significantly increasing mobility. The density of twin boundaries, which also limit mobility, could be reduced through the use of miscut substrates[20,33,40]. The high residual carrier density may be caused by an accumulation layer that arises at the film/substrate interface and conducts in parallel with the Bi$_2$Se$_3$ film, which has been predicted by some calculations.[41] This accumulation layer could be controlled through the use of a back gate or by using an alternative semiconductor as a top layer in the heterostructure to engineer the band lineup.

One additional question concerns the GaAs termination at the interface, especially in the case of the SL sample with the atomically smooth interface. Normally, a GaAs substrate deoxidized under an arsenic flux would be arsenic terminated. However, for all three films, the substrate is subjected to a selenium flux (and a bismuth flux) upon the initiation of the growth of the Bi$_2$Se$_3$ film. This exposure to the selenium flux could alter the GaAs substrate termination. Understanding the details of the interface is critical to understanding the transport and optical properties of the structure. We, therefore, conducted density functional theory (DFT) modeling of the Bi$_2$Se$_3$/GaAs interface structure. The lattice structure of the interface model is shown in **Figure 3**. The interface formation energy as a function of the As and Se chemical potentials was calculated using the expression:

$$E_f[\text{Bi}_2\text{Se}_3/\text{GaAs}] = E_{tot}[\text{Bi}_2\text{Se}_3/\text{GaAs}] - E_{tot}(\text{Bi}_2\text{Se}_3) - E_{tot}(\text{GaAs}) + \sum n_i \mu_i \ (1.)$$

where $E_{tot}(\text{Bi}_2\text{Se}_3/\text{GaAs})$ is the total energy of the supercell for the Bi$_2$Se$_3$ (0001)$_H$/GaAs (001)$_C$ interface, $E_{tot}(\text{Bi}_2\text{Se}_3)$ is the total energy of the Bi$_2$Se$_3$ slab, and $E_{tot}(\text{GaAs})$ is the total energy of the (1x1) non-relaxed, ideal As-terminated GaAs (001)$_C$ slab with the bottom interface being terminated by Ga passivated with H atoms with a fictitious charge of 1.25; $n_i$ is the number of atoms of type *i* removed or added to the supercell to form the interface, and $\mu_i$ is the corresponding atomic chemical potential of atom type *i*. Thus, the formation energy is taken with respect the As-terminated GaAs (001)$_C$ interface, i.e., $n_i = 0$. In the case of the Se-terminated interface, the topmost atomic layer of As atoms is replaced with Se atoms. The number of As atoms replaced by Se atoms at the interface represents the Se coverage of the GaAs (001)$_C$ substrate. We have considered Se coverages of ~43%, ~57%, ~71%, and 100%, corresponding to replacing 3/7, 4/7, 5/7, and 7/7 As with Se atoms in the supercell (see Computational Approach in Methods). The chemical potentials $\mu_{As}$ and $\mu_{Se}$ are treated as variables to represent different growth or processing conditions and are referenced to the total energy per atom of As and Se bulk phases, i.e., $E_{tot}(\text{As}_{bulk})$ and $E_{tot}(\text{Se}_{bulk})$, repsectively. These chemical potentials satisfy the stability condition of GaAs and Bi$_2$Se$_3$, i.e.,

$$2\mu_{Bi} + 3\mu_{Se} = \Delta H_f(\text{Bi}_2\text{Se}_3) \quad (2.)$$

$$\mu_{Ga} + \mu_{As} = \Delta H_f(\text{GaAs}) \quad (3.)$$

where $\Delta H_f(\text{Bi}_2\text{Se}_3) = -2.18 \text{ eV/f.u.}$ and $\Delta H_f(\text{GaAs}) = -0.64 \text{ eV/f.u.}$ are the enthalpies of formation of Bi$_2$Se$_3$ and GaAs, and $\mu_{As}, \mu_{Se} \leq 0$. Therefore, the chemical potentials $\mu_{As}$ and $\mu_{Se}$ vary in in the following ranges:

$$\Delta H_f(\text{GaAs}) \leq \mu_{As} \leq 0 \quad (4.)$$

$$\frac{1}{3}\Delta H_f(\text{Bi}_2\text{Se}_3) \leq \mu_{Se} \leq 0 \quad (5.)$$

with $\mu_{As} = 0$ and $\mu_{Se} = 0$ corresponding to the As-rich and Se-rich limit.

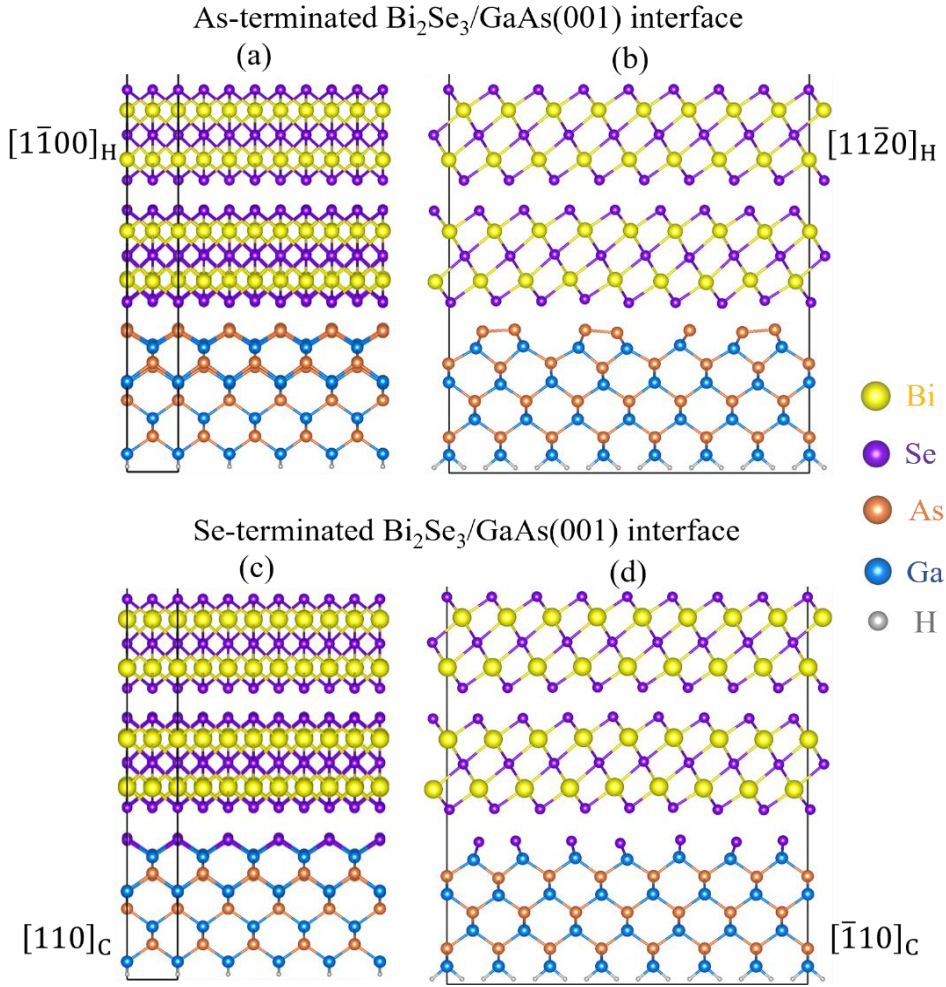

As-terminated Bi$_2$Se$_3$/GaAs(001) interface
(a) (b)

Se-terminated Bi$_2$Se$_3$/GaAs(001) interface
(c) (d)

**Figure 3.** Supercells used to simulate the Bi$_2$Se$_3$/GaAs (0001)$_H$ interface, with vacuum added above the structure. The (relaxed) As-terminated interface is shown in (a) and (b), and the Se-terminated interface (100% Se coverage is shown in (c) and (d). The view directions are [110]$_C$ for GaAs and [1$\bar{1}$00]$_H$ for Bi$_2$Se$_3$ in (a) and (c), and [$\bar{1}$10]$_C$ for GaAs and [11$\bar{2}$0]$_H$ for Bi$_2$Se$_3$ in (b) and (d).

Thus, the interface formation energy for a given Se coverage larger than zero will depend on the $\mu_{As}$ and $\mu_{Se}$ chemical potentials according to Eq. 1. The calculated interface formation energies as a function of Se coverage at two limiting conditions (As-rich/Se-poor and As-poor/Se-rich) are listed in **Table 1**. First, for the As-terminated interface, the interface formation energy does not depend on $\mu_{As}$ and $\mu_{Se}$ since $n_{Se} = n_{As} = 0$, and the interface formation energy represents the energy gained by forming an interface from a slab of Bi$_2$Se$_3$ and an As-terminated GaAs slab. Second, our results show that the interfaces with high Se coverages have lower formation energies. We find that the lowest interface energy is obtained for 100% Se-covered GaAs at the Bi$_2$Se$_3$ (0001)$_H$/GaAs (001)$_C$ interface in the As-poor/Se-rich limit condition; this configuration is 97 meV/Å$^2$ lower in energy than the As-terminated interface. One can understand the stability of the Se-terminated Bi$_2$Se$_3$/GaAs (001)$_C$ and its structure as follows: in the case of the As-terminated GaAs (001)$_C$, the As dangling bonds are partially occupied with 5/4 electrons each, thus tending to form As-As dimers as indicated in Figure 3(b). By replacing the top As monolayer with Se, each Se dangling bond will contain 7/4 electrons.[42] This almost completely filled dangling bonds situation precludes the formation of Se-Se dimers in the Se-terminated Bi$_2$Se$_3$/GaAs (001)$_C$ interface, as shown in Figure 3(c) and (d), and consistent with the fact that selenization of the GaAs (001)$_C$ interface is performed before Bi$_2$Se$_3$ is deposited. These almost filled Se dangling bonds also give rise to partially occupied interface states that, in principle, contribute to the transport measurements of the heterostructure discussed previously. Therefore, the presence of a Ga-Se layer must be considered when characterizing the optical and transport properties of Bi$_2$Se$_3$/GaAs structures grown using the conditions described above.

**Table 1.** Interface formation energies (in meV/Å$^2$) of the Bi$_2$Se$_3$/GaAs (001)$_C$ interface at various Se-coverage of the GaAs layer.

| % Se coverage | 0% /As-terminated | ~43% | ~57% | ~71% | 100% |
| --- | --- | --- | --- | --- | --- |
| As-rich/Se-poor | -30 | -37 | -46 | -45 | -42 |
| As-poor/Se-rich | -30 | -73 | -95 | -106 | -127 |

To inspect the nature of the bonding across the Se-terminated $Bi_2Se_3$/GaAs interface, we computed the electron localization function (ELF)[43-45], which can take values between 0 and 1. ELF can be used to classify bonds of different types, i.e., chemical bonding such as covalent or ionic bonds, or physical bonding through the van der Waals interaction [45,46]. High values of ELF (i.e., close to 1) indicate a high degree of localization of electron at a given position. Highly localized electron density regions usually correspond to covalent bonds, core electrons, or lone pair electron regions around an atom. Small values (i.e., close to 0) can be assigned to regions of delocalized (itinerant) electrons, typically found in noncovalent bonding environments such as metallic systems, ionic bonding regions, or van der Waals gaps. As a reference, ELF = 0.5 corresponds to a homogeneous electron gas system. We show in **Figure 4** the computed ELF in the plane containing Se atoms of the Se-terminated GaAs (001)$_C$ (*cf.* Figure 3(d)) and the Se atoms from the bottom of $Bi_2Se_3$ layer of the heterointerface. It is evident from the ELF plot that there is no Se-Se covalent bonding or any build-up of charge along the lines connecting two neighboring Se atoms across the interface, indicating that the interaction between the Se-terminated GaAs and $Bi_2Se_3$ is of van der Waals type. This is corroborated by the fact that the distances between two neighboring Se atoms across the interface are larger than 2.83 Å and much longer than twice the covalent radius of Se atoms of 2.32 Å, thus reinforcing our conclusion of the van der Waals type of interaction at the interface.

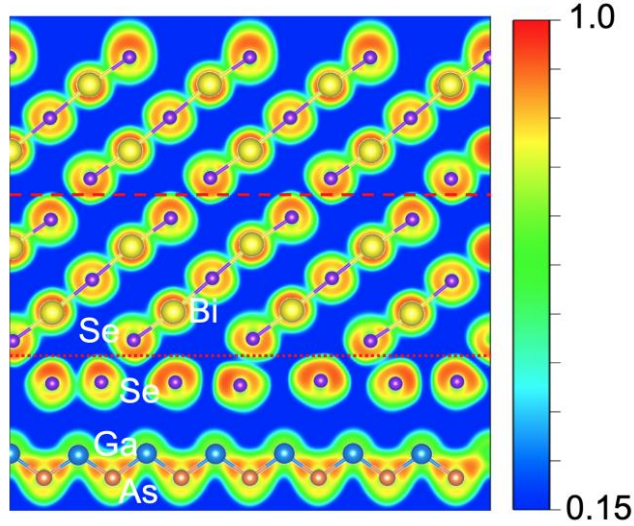

**Figure 4.** Electron localization function (ELF) plotted on the $(11\bar{2}0)_H$ plane containing Se atoms across the interface in the $Bi_2Se_3$/GaAs $(001)_C$ heterostructure showing no covalent bonding between the Se-terminated GaAs $(001)_C$ substrate and $Bi_2Se_3$. The red dashed line and red dotted line indicate the van der Waals gap regions between $Bi_2Se_3$ quintuple layers and interface between GaAs and $Bi_2Se_3$, respectively.

**Conclusion**

In summary, we have shown that the TI vdW material $Bi_2Se_3$ can be grown as a single-crystalline single-orientation film with low surface roughness on semiconductor substrates with appropriate substrate pre-treatment conditions. In particular, atomically flat substrates are needed to grow smooth, single-crystalline films, and this condition can be achieved through the use of smoothing superlattices after thermal oxide desorption. Using DFT calculations, we determine that the interface is most likely selenium-terminated with a van der Waals gap between the film and substrate. This conclusion is supported by STEM imaging. Understanding how to grow smooth, single-orientation $Bi_2Se_3$ films on technologically-relevant III-V materials and understanding the chemical nature of the interface between them is critical for the creation of a THz integrated system design. More broadly, the ability to integrate van der Waals materials with good crystal quality on common, commercially-available substrates is necessary if we wish to create next-generation optoelectronic devices using these materials. This result is the first step toward that realization.

**Methods**

**Epitaxial Thin Film Growth**

The Bi$_2$Se$_3$ thin films were synthesized on (100)$_C$-oriented single-side polished GaAs 2-inch quarter wafer substrates using molecular beam epitaxy (MBE). Samples were grown in the University of Delaware Materials Growth Facility using two Veeco GenXplor MBE systems connected by an ultra-high vacuum transfer tube. One system is dedicated to the growth of III-V compounds while the other is used for chalcogenide materials. The three samples underwent different treatments in the III-V MBE. For Se-desorbed sample, the process only consisted of the transfer from the III-V MBE system to the TI MBE system without any treatment in the III-V MBE. For As-desorbed sample, the native oxide from the semi-insulating GaAs (001)$_C$ wafer was desorbed in the III-V MBE system by thermal deoxidation with an As$_2$ overpressure. The beam equivalent pressure of As$_2$ is $1 \times 10^{-5}$ Torr. After cooling the sample, it was transferred in an ultra-high vacuum environment from the III-V MBE system to the TI MBE system for Bi$_2$Se$_3$ film growth as described below. For SL sample, the native oxide from the wafer was desorbed in the III-V MBE system using thermal deoxidation as described above. After the oxide desorption, the temperature was decreased to 580°C to grow a 100 nm GaAs/AlAs smoothing superlattice with a 10 nm period. A 50 nm GaAs buffer layer was grown on top using the same conditions. After cooling the sample, it was transferred in an ultra-high vacuum environment from the III-V MBE system to the TI MBE system for Bi$_2$Se$_3$ film growth as described below. All substrate temperatures are measured by band edge thermometry (BET).

After entering the chalcogenide system, Se-desorbed sample was heated to 650°C measured by a noncontact thermocouple for thermal deoxidation. A selenium flux is applied during this process to prevent gallium droplet formation. After deoxidation, the substrate is cooled to 350°C for the subsequent Bi$_2$Se$_3$ film growth. For SL sample and As-desorbed sample, the wafers were directly loaded into the growth chamber of the chalcogenide MBE for the subsequent $Bi_2Se_3$ film growth at 350°C. All the growths used a recipe of 1 minute growth and 1 minute anneal. The growth rate is 0.74 nm/minute, and the film thickness is 50 nm. A dual-filament Knudson effusion cell is used to generate the bismuth flux while a cracker source is used for selenium to ensure efficient incorporation. An ion gauge in the substrate position is used to monitor the fluxes. During growth, the film quality is monitored by an *in situ* reflection high energy electron diffraction (RHEED) system. After growth, the samples were removed from the chalcogenide MBE and sealed in a vacuum pack to await further characterization.

**Scanning Transmission Electron Microscopy**

An FEI Nova NanoLab 600 DualBeam (SEM/FIB) was employed to prepare cross-sectional STEM samples. Platinum was initially deposited on top of the films to protect the sample surface using electron beam deposition. To reduce the damage from the gallium ions, in the final step of preparation the STEM samples were thinned with 2 kV Ga-ions using a low beam current of 29 pA. An FEI Titan 80-300 probe-corrected STEM/TEM microscope operating at 300 keV was employed to acquire atomic-resolution high-angle annular dark field (HAADF) STEM images.

**X-Ray Diffraction**

A Rigaku Ultima IV XRD system was used for x-ray diffraction measurements. Spectra were taken separately for different crystallographic directions (parallel and perpendicular to the GaAs $[0\bar{1}\bar{1}]_C$ main flat). A Cu K$\alpha$1 source (1.504 eV) was used for the x-ray generation. Data were taken with 2θ ranging from 15° to 50° with 0.02° step size and 0.5 deg/min.

**Atomic Force Microscopy**

Samples were characterized with the Dimension 3100 (D3100) atomic force microscope under normal atmosphere and at room temperature. The sample scan size is 5 μm x 5 μm using tapping mode.

**X-ray Pole Scans**

Pole scan characterization was conducted using the Rigaku Ultima IV XRD system. This technique is an in-plane measurement. After putting samples into the system and initializing, specimen alignment and precise alignment were used to align the Z, ω and 2θ parameters (0.5 mm split). Phi scans from 0° to 360° were taken with speed 40.00 °/min, 0.050° sampling step, ATT open, DS 1.00 mm, SS 0.5 mm and RS 0.50 mm.

**Computational Approach**

The calculations are based on the density functional theory (DFT) [47,48] within the semilocal PBEsol approximation,[49] projector augmented wave (PAW) potentials,[50] and plane-wave basis set as implemented in the Vienna Ab initio Simulation Package (VASP).[51,52] Dispersion or van der Waals interactions are included using the D3 method of Grimme *et al*.[53] Plane-wave cutoff of 400 eV was employed with a Γ-centered mesh of 7 x 1 x 1 used to sample the Brillouin zone of the slab supercell representing the $Bi_2Se_3$ $(0001)_H$/GaAs $(001)_C$ interface.

The modelled interface supercell contains two quintuple layers of $Bi_2Se_3$ and four layers of GaAs bilayers with the Ga-terminated bottom layer passivated by artificial H atoms with charge of 1.25 electrons, as shown in Figure 3. To accommodate the lattice mismatch between $Bi_2Se_3$ $(0001)_H$ and GaAs $(001)_C$, we strained the $Bi_2Se_3$ layer by 3.4% in the $[11\bar{2}0]_H$ direction and by 2.4% in the $[1\bar{1}00]_H$ direction. On the top surface of GaAs $(001)_C$ slab there are 7 As atoms in the supercell. A vacuum spacing of over 15 Å was employed to minimize the spurious interaction between slabs along the direction perpendicular to the interface. All atoms, except the bottom Ga-H layers were allowed to relax until the residual force on each of them was less than 0.01 eV/Å.

Different terminations for the $Bi_2Se_3$/GaAs (001)$_C$ interface were explored, namely, Ga-, As-, and Se-terminated GaAs (001)$_C$.

**Data availability** Data is available upon reasonable request of the authors.

**Supporting Information** Experimental x-ray diffraction pole scans and analysis of all three samples to determine degree of twinning. Samples show similar degrees of twinning (almost 1:1), with a small improvement in the SL sample.

**Acknowledgements**

This research was primarily supported by the NSF through the University of Delaware Materials Research Science and Engineering Center (No. DMR-2011824). The authors acknowledge the use of the Materials Growth Facility (MGF) at the University of Delaware, which is partially supported by the National Science Foundation Major Research Instrumentation under Grant No. 1828141 and UD-CHARM, a National Science Foundation MRSEC, under Award No. DMR-2011824. Product names are mentioned to provide an accurate record of what was done. Reference of product names does not constitute validation or endorsement. H.Z. acknowledges support from the U.S. Department of Commerce, National Institute of Standards and Technology under the financial assistance awards 70NANB19H138. A.V.D. acknowledges the support of Material Genome Initiative funding allocated to NIST.

**Author contributions** Y.L. and W.A. synthesized the samples. H.Z. performed the STEM measurements. Y.L. and Z.W. performed the AFM, XRD, and Hall effect measurements. D.Q.H. and R.H. performed DFT calculations. A.J., G.B., A.V.D., J.M.O.Z., and S.L. designed the scientific objectives, oversaw the project, and co-wrote the manuscript. All authors discussed the results and commented on the manuscript.

**Conflict of interest:** The authors declare no conflict of interest.

**Reference**

[1] Fitzgerald, A. J.; Berry, E.; Zinovev, N. N.; Walker, G. C.; Smith, M. A.; Chamberlain, J. M. An Introduction to Medical Imaging with Coherent Terahertz Frequency Radiation. Physics in Medicine and Biology 2002, 47 (7).

[2] Liu, H.-B.; Zhong, H.; Karpowicz, N.; Chen, Y.; Zhang, X.-C. Terahertz Spectroscopy and Imaging for Defense and Security Applications. Proceedings of the IEEE 2007, 95 (8), 1514–1527.

[3] D'Arco, A.; Di Fabrizio, M. D.; Dolci, V.; Petrarca, M.; Lupi, S. THz Pulsed Imaging in Biomedical Applications. Condensed Matter 2020, 5 (2), 25.

[4] Heydari, P. Terahertz Integrated Circuits and Systems for High-Speed Wireless Communications: Challenges and Design Perspectives. IEEE Open Journal of the Solid-State Circuits Society 2021, 1, 18–36.

[5] Afsah-Hejri, L.; Hajeb, P.; Ara, P.; Ehsani, R. J. A Comprehensive Review on Food Applications of Terahertz Spectroscopy and Imaging. Comprehensive Reviews in Food Science and Food Safety 2019, 18 (5), 1563–1621.

[6] Zhang, Y.; Wang, C.; Huai, B.; Wang, S.; Zhang, Y.; Wang, D.; Rong, L.; Zheng, Y. Continuous-Wave THz Imaging for Biomedical Samples. Applied Sciences 2020, 11 (1), 71.

[7] Sizov, F.; Rogalski, A. THz Detectors. Progress in Quantum Electronics 2010, 34 (5), 278–347.

[8] Ma, Z. T.; Geng, Z. X.; Fan, Z. Y.; Liu, J.; Chen, H. D. Modulators for Terahertz Communication: The Current State of the Art. Research 2019, 2019, 1–22.

[9] Singh, K.; Kumar, S.; Bandyopadhyay, A.; Sengupta, A. Characterization of Hollow-Core-Metal Waveguide Using Broadband THz Time Domain Spectroscopy for High-Pressure and Temperature Sensor. Terahertz, RF, Millimeter, and Submillimeter-Wave Technology and Applications XIV 2021.

[10] Nasir, S.; Wang, Z.; Mambakkam, S. V.; Law, S. In-Plane Plasmon Coupling in Topological Insulator $Bi_2Se_3$ Thin Films. Applied Physics Letters 2021, 119 (20), 201103.

[11] Wang, Z.; Ginley, T. P.; Mambakkam, S. V.; Chandan, G.; Zhang, Y.; Ni, C.; Law, S. Plasmon Coupling in Topological Insulator Multilayers. Physical Review Materials 2020, 4 (11).

[12] Di Pietro, P.; Adhlakha, N.; Piccirilli, F.; Di Gaspare, A.; Moon, J.; Oh, S.; Di Mitri, S.; Spampinati, S.; Perucchi, A.; Lupi, S. Terahertz tuning of Dirac plasmons in $Bi_2Se_3$ topological insulator. Physical Review Letters 2020, 124(22).

[13] Autore, M.; D'Apuzzo, F.; Di Gaspare, A.; Giliberti, V.; Limaj, O.; Roy, P.; Brahlek, M.; Koirala, N.; Oh, S.; García de Abajo, F. J.; Lupi, S. Plasmon-Phonon Interactions in Topological Insulator Microrings. Advanced Optical Materials 2015, 3 (9), 1257–1263.

[14] Ginley, T. P.; Zhang, Y.; Ni, C.; Law, S. Epitaxial Growth of $Bi_2Se_3$ in the (0015) Orientation on GaAs (001). Journal of Vacuum Science & Technology A 2020, 38 (2), 023404.

[15] Wang, Z.; Law, S. Optimization of the Growth of the Van Der Waals Materials $Bi_2Se_3$ and $(Bi_{0.5}In_{0.5})_2Se_3$ by Molecular Beam Epitaxy. Crystal Growth & Design 2021, 21 (12), 6752–6765.

[16] Ginley, T. P.; Law, S. Growth of $Bi_2Se_3$ Topological Insulator Films Using a Selenium Cracker Source. Journal of Vacuum Science & Technology B, Nanotechnology and Microelectronics: Materials, Processing, Measurement, and Phenomena 2016, 34 (2).

[17] Levy, I.; Garcia, T. A.; Shafique, S.; Tamargo, M. C. Reduced Twinning and Surface Roughness of $Bi_2Se_3$ and $Bi_2Te_3$ Layers Grown by Molecular Beam Epitaxy on Sapphire Substrates. Journal of Vacuum Science & Technology B, Nanotechnology and Microelectronics: Materials, Processing, Measurement, and Phenomena 2018, 36 (2), 02D107.

[18] Geižutis, A.; Adomavičius, R.; Urbanowicz, A.; Bertulis, K.; Krotkus, A.; Tan, H. H.; Jagadish, C. Carrier Recombination Properties in Low-Temperature-Grown and Ion-Implanted GaAs. Lithuanian Journal of Physics 2005, 45 (4), 249–255.

[19] O'Hara, J. F.; Zide, J. M.; Gossard, A. C.; Taylor, A. J.; Averitt, R. D. Enhanced Terahertz Detection via ErAs:GaAs Nanoisland Superlattices. Applied Physics Letters 2006, 88 (25), 251119.

[20] Chen, Z.; Garcia, T. A.; De Jesus, J.; Zhao, L.; Deng, H.; Secor, J.; Begliarbekov, M.; Krusin-Elbaum, L.; Tamargo, M. C. Molecular Beam Epitaxial Growth and Properties of $Bi_2Se_3$ Topological Insulator Layers on Different Substrate Surfaces. Journal of Electronic Materials 2013, 43 (4), 909–913.

[21] Richardella, A.; Zhang, D. M.; Lee, J. S.; Koser, A.; Rench, D. W.; Yeats, A. L.; Buckley, B. B.; Awschalom, D. D.; Samarth, N. Coherent Heteroepitaxy of $Bi_2Se_3$ on GaAs (111)B. Applied Physics Letters 2010, 97 (26), 262104.

[22] Hagmann, J. A.; Li, X.; Chowdhury, S.; Dong, S.-N.; Rouvimov, S.; Pookpanratana, S. J.; Man Yu, K.; Orlova, T. A.; Bolin, T. B.; Segre, C. U.; Seiler, D. G.; Richter, C. A.; Liu, X.; Dobrowolska, M.; Furdyna, J. K. Molecular Beam Epitaxy Growth and Structure of Self-Assembled $Bi_2Se_3$/$Bi_2MnSe_4$ Multilayer Heterostructures. New Journal of Physics 2017, 19 (8), 085002.

[23] Eddrief, M.; Atkinson, P.; Etgens, V.; Jusserand, B. Low-Temperature Raman Fingerprints for Few-Quintuple Layer Topological Insulator $Bi_2Se_3$ Films Epitaxied on GaAs. Nanotechnology 2014, 25 (24), 245701.

[24] Liu, X.; Smith, D. J.; Cao, H.; Chen, Y. P.; Fan, J.; Zhang, Y.-H.; Pimpinella, R. E.; Dobrowolska, M.; Furdyna, J. K. Characterization of $Bi_2Te_3$ and $Bi_2Se_3$ Topological Insulators Grown by MBE on (001) GaAs Substrates. Journal of Vacuum Science &

Technology B, Nanotechnology and Microelectronics: Materials, Processing, Measurement, and Phenomena 2012, 30 (2), 02B103.

[25] Liu, X.; Smith, D. J.; Fan, J.; Zhang, Y.-H.; Cao, H.; Chen, Y. P.; Leiner, J.; Kirby, B. J.; Dobrowolska, M.; Furdyna, J. K. Structural Properties of $Bi_2Te_3$ and $Bi_2Se_3$ Topological Insulators Grown by Molecular Beam Epitaxy on GaAs(001) Substrates. Applied Physics Letters 2011, 99 (17), 171903.

[26] Li, B.; Chen, W.; Guo, X.; Ho, W.; Dai, X.; Jia, J.; Xie, M. Strain in Epitaxial High-Index $Bi_2Se_3$(221) Films Grown by Molecular-Beam Epitaxy. Applied Surface Science 2017, 396, 1825–1830.

[27] Petroff, P. M.; Miller, R. C.; Gossard, A. C.; Wiegmann, W. Impurity Trapping, Interface Structure, and Luminescence of GaAs Quantum Wells Grown by Molecular Beam Epitaxy. Applied Physics Letters 1984, 44 (2), 217–219.

[28] Drummond, T. J.; Klem, J.; Arnold, D.; Fischer, R.; Thorne, R. E.; Lyons, W. G.; Morkoç, H. Use of a Superlattice to Enhance the Interface Properties between Two Bulk Heterolayers. Applied Physics Letters 1983, 42 (7), 615–617.

[29] Petroff, P. M.; Gossard, A. C.; Wiegmann, W.; Savage, A. Crystal Growth Kinetics in $(GaAs)_n-(AlAs)_m$ Superlattices Deposited by Molecular Beam Epitaxy. Journal of Crystal Growth 1978, 44 (1), 5–13.

[30] Weisbuch, C.; Dingle, R.; Petroff, P. M.; Gossard, A. C.; Wiegmann, W. Dependence of the Structural and Optical Properties of GaAs-$Ga_{1-x}Al_xAs$ Multiquantum-Well Structures on Growth Temperature. Applied Physics Letters 1981, 38 (11), 840–842.

[31] Tarakina, N. V.; Schreyeck, S.; Luysberg, M.; Grauer, S.; Schumacher, C.; Karczewski, G.; Brunner, K.; Gould, C.; Buhmann, H.; Dunin-Borkowski, R. E.; Molenkamp, L. W. Suppressing Twin Formation in $Bi_2Se_3$ Thin Films. Advanced Materials Interfaces 2014, 1 (5), 1400134.

[32] Xu, Z.; Guo, X.; Yao, M.; He, H.; Miao, L.; Jiao, L.; Liu, H.; Wang, J.; Qian, D.; Jia, J.; Ho, W.; Xie, M. Anisotropic Topological Surface States on High-Index $Bi_2Se_3$ Films. Advanced Materials 2013, 25 (11), 1557–1562.

[33] Schreyeck, S.; Tarakina, N. V.; Karczewski, G.; Schumacher, C.; Borzenko, T.; Brüne, C.; Buhmann, H.; Gould, C.; Brunner, K.; Molenkamp, L. W. Molecular Beam Epitaxy of High Structural Quality $Bi_2Se_3$ on Lattice Matched InP(111) Substrates. Applied Physics Letters 2013, 102 (4), 041914.

[34] Tarakina, N. V.; Schreyeck, S.; Borzenko, T.; Schumacher, C.; Karczewski, G.; Brunner, K.; Gould, C.; Buhmann, H.; Molenkamp, L. W. Comparative Study of the Microstructure of $Bi_2Se_3$ Thin Films Grown on Si(111) and InP(111) Substrates. Crystal Growth & Design 2012, 12 (4), 1913–1918.

[35] van der PAUW, L. J. A Method of Measuring Specific Resistivity and Hall Effect of Discs of Arbitrary Shape. Semiconductor Devices: Pioneering Papers 1991, 174–182.

[36] Bianchi, M.; Guan, D.; Bao, S.; Mi, J.; Iversen, B. B.; King, P. D. C.; Hofmann, P. Coexistence of the Topological State and a Two-Dimensional Electron Gas on the Surface of $Bi_2Se_3$. Nature Communications 2010, 1 (1), 128.

[37] Koirala, N.; Brahlek, M.; Salehi, M.; Wu, L.; Dai, J.; Waugh, J.; Nummy, T.; Han, M.-G.; Moon, J.; Zhu, Y.; Dessau, D.; Wu, W.; Armitage, N. P.; Oh, S. Record Surface State Mobility and Quantum Hall Effect in Topological Insulator Thin Films via Interface Engineering. Nano Letters 2015, 15 (12), 8245–8249.

[38] Yao, X.; Moon, J.; Cheong, S.-W.; Oh, S. Structurally and Chemically Compatible $BiInSe_3$ Substrate for Topological Insulator Thin Films. Nano Research 2020, 13 (9), 2541–2545.

[39] Wang, Y.; Ginley, T. P.; Law, S. Growth of High-Quality $Bi_2Se_3$ Topological Insulators Using $(Bi_{1-x}In_x)_2Se_3$ Buffer Layers. Journal of Vacuum Science & Technology B, Nanotechnology and Microelectronics: Materials, Processing, Measurement, and Phenomena 2018, 36 (2), 02D101.

[40] Takagaki, Y.; Jenichen, B.; Tominaga, J. Giant Corrugations in $Bi_2Se_3$ Layers Grown on High-Index InP Substrates. Physical Review B 87, 245302.

[41] Seixas, L.; West, D.; Fazzio, A.; Zhang, S. B. Vertical Twinning of the Dirac Cone at the Interface between Topological Insulators and Semiconductors. Nature Communications 2015, 6 (1), 7630.

[42] Farrell, H. H.; Tamargo, M. C.; de Miguel, J. L.; Turco, F. S.; Hwang, D. M.; Nahory, R. E. "Designer" Interfaces in II-VI/III-V Polar Heteroepitaxy. Journal of Applied Physics 1991, 69 (10), 7021–7028.

[43] Becke, A. D.; Edgecombe, K. E. A Simple Measure of Electron Localization in Atomic and Molecular Systems. The Journal of Chemical Physics 1990, 92 (9), 5397–5403.

[44] Savin, A.; Jepsen, O.; Flad, J.; Andersen, O. K.; Preuss, H.; von Schnering, H. G. Electron Localization in Solid-State Structures of the Elements: The Diamond Structure. Angewandte Chemie International Edition in English 1992, 31 (2), 187–188.

[45] Silvi, B.; Savin, A. Classification of Chemical Bonds Based on Topological Analysis of Electron Localization Functions. Nature 1994, 371 (6499), 683–686.

[46] Koumpouras, K.; Larsson, J. A. Distinguishing between Chemical Bonding and Physical Binding Using Electron Localization Function (ELF). Journal of Physics: Condensed Matter 2020, 32 (31), 315502.

[47] Hohenberg, P.; Kohn, W. Inhomogeneous Electron Gas. Physical Review 1964, 136 (3B).

[48] Kohn, W.; Sham, L. J. Self-Consistent Equations Including Exchange and Correlation Effects. Physical Review 1965, 140 (4A).

[49] Perdew, J. P.; Ruzsinszky, A.; Csonka, G. I.; Vydrov, O. A.; Scuseria, G. E.; Constantin, L. A.; Zhou, X.; Burke, K. Restoring the Density-Gradient Expansion for Exchange in Solids and Surfaces. Physical Review Letters 2008, 100 (13).

[50] Blöchl, P. E. Projector Augmented-Wave Method. Physical Review B 1994, 50 (24), 17953–17979.

[51] Kresse, G.; Furthmüller, J. Efficient Iterative Schemes for Ab-Initio Total Energy Calculations Using a Plane-Wave Basis Set. Physical Review B 1996, 54 (16), 11169–11186.

[52] Kresse, G.; Joubert, D. From Ultrasoft Pseudopotentials to the Projector Augmented-Wave Method. Physical Review B 1999, 59 (3), 1758–1775.

[53] Grimme, S.; Antony, J.; Ehrlich, S.; Krieg, H. A Consistent and Accurate Ab-Initio Parametrization of Density Functional Dispersion Correction (DFT-D) for the 94 Elements H-Pu. The Journal of Chemical Physics 2010, 132 (15), 154104.

# Supporting Information

# Bi$_2$Se$_3$ Growth on (001) GaAs Substrates for Terahertz Integrated Systems

Yongchen Liu[1], Wilder Acuna[1], Huairuo Zhang[2,3], Dai Q. Ho[1], Ruiqi Hu[1], Zhengtianye Wang[1], Anderson Janotti[1], Garnett Bryant[4], Albert V. Davydov[2], Joshua M. O. Zide[1], and Stephanie Law[1]*

[1] Department of Materials Science and Engineering, University of Delaware, Newark DE 19716 USA
[2] Materials Science and Engineering Division, National Institute of Standards and Technology, Gaithersburg, MD 20899 USA
[4] Nanoscale Device Characterization Division, National Institute of Standards and Technology, Gaithersburg, MD 20899 USA
[3] Theiss Research, Inc., La Jolla, CA 92037 USA

*E-mail: slaw@udel.edu

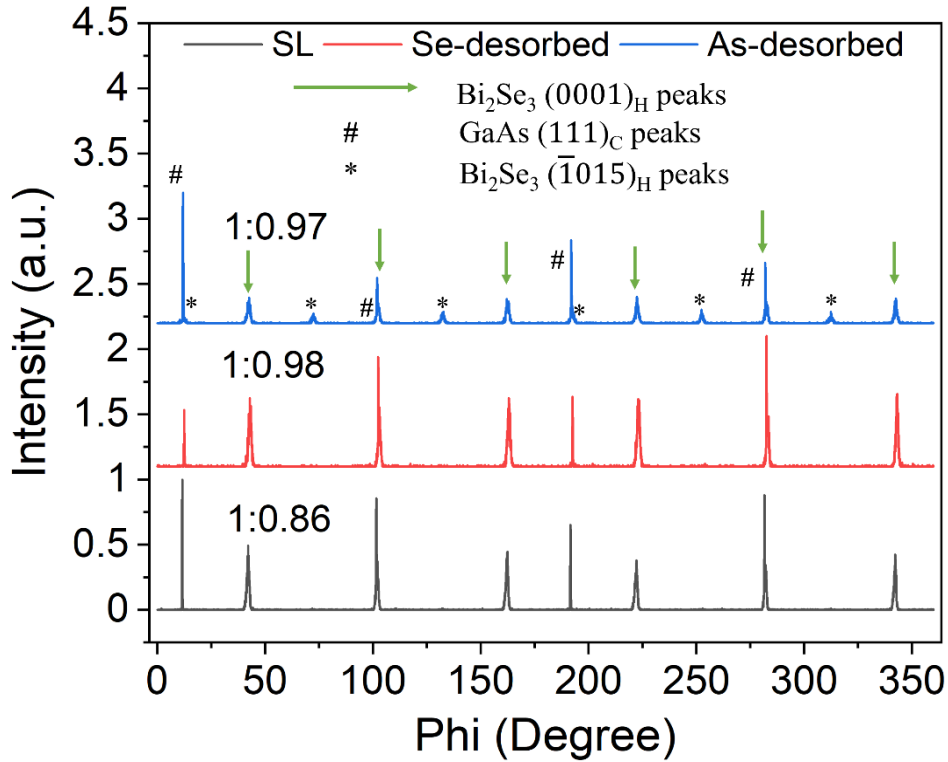

**Figure S1.** Pole scan XRD spectra. The ratios above the spectra are the twining ratio of the samples. Arrows show the peaks for $Bi_2Se_3$ $(0001)_H$. # symbols indicate the GaAs $(111)_C$ peaks and * symbols indicate the $(\bar{1}015)_H$ $Bi_2Se_3$ peaks.

From the AFM results, we can see two sets of triangular domains rotated 60º with respect to each other. These are commonly called twin defects or antiphase domains, and they are caused by the in-plane six-fold symmetry of the $Bi_2Se_3$ lattice [1]. The film can nucleate in two different orientations, causing grain boundaries when the domains coalesce. These defects reduce the charge carrier mobility of the film since they act as scattering centers. To quantify the degree of twinning in the films, we performed x-ray pole scans around the $(\bar{1}015)_H$ reflection, shown in **Figure S1**. We label the peaks with green arrows ($Bi_2Se_3$ $(0001)_H$ orientation), # (GaAs $(111)_C$ orientation), or * ($Bi_2Se_3$ $(\bar{1}015)_H$ orientation). We see peaks arising from the $Bi_2Se_3$ $(\bar{1}015)_H$ orientation only in the Se-desorbed and As-desorbed samples, consistent with the XRD measurements. We can quantify the degree of twinning by comparing the intensity ratio of the peaks associated with one domain (near 40º, 160º, and 280º) to the peaks associated with the other (near 100º, 220º, and 340º). In this case, the GaAs substrate peaks overlap with the $Bi_2Se_3$ peaks at 100º and 280º, so the other four peaks were used. All three samples showed a similar degree of twinning, though the SL sample is slightly better. This is not surprising, as the different surface treatments were only intended to control the substrate roughness, not the surface energy. Both domain orientations will therefore continue to nucleate with roughly equal probability.

**Reference**

[1] Tarakina, N. V.; Schreyeck, S.; Luysberg, M.; Grauer, S.; Schumacher, C.; Karczewski, G.; Brunner, K.; Gould, C.; Buhmann, H.; Dunin-Borkowski, R. E.; Molenkamp, L. W. Suppressing Twin Formation in $Bi_2Se_3$ Thin Films. Advanced Materials Interfaces 2014, 1 (5), 1400134.